\documentclass[twocolumn,aps,prd,eqsecnum,nofootinbib]{revtex4}
\usepackage{epsfig}
\usepackage{amssymb}
\usepackage{amsmath}
\usepackage{amsfonts}
\usepackage{bm}

\newcommand{\be}{\begin{equation}}
\newcommand{\ee}{\end{equation}}
\newcommand{\bea}{\begin{eqnarray}}
\newcommand{\eea}{\end{eqnarray}}

\DeclareMathSymbol{\R}{\mathbin}{AMSb}{"52}
\begin{document}

\title{Detectability of Mode Resonances in Coalescing Neutron Star Binaries}
\author{Prakash Balachandran and \'Eanna \'E. Flanagan}
\affiliation{Laboratory for Elementary Particle Physics, Cornell University, Ithaca, NY 14853, USA.}
\date{\today}

\begin{abstract}
Inspirals of neutron star-neutron star binaries are a
promising source of gravitational waves for gravitational wave
detectors like LIGO.  During the
inspiral, the tidal gravitational
field of one of the stars can resonantly excite internal modes of the
other star, resulting in a
phase shift in the gravitational wave
signal.  We compute using a Fisher-matrix analysis how large the phase
shift must be in order to be detectable.  For a $1.4 M_\odot, 1.4
M_\odot$ binary the result is $\sim 8.1, 2.9$ and $1.8$ radians, for resonant
frequencies of $16, 32$ and $64$ Hz.  The measurement accuracies of the other
binary parameters are degraded by inclusion of the mode resonance
effect.


\end{abstract}
\maketitle

\section{Introduction and summary}

In neutron star - neutron star (NS-NS)
binaries, each star exerts a tidal gravitational
force on its companion as the stars inspiral.
This force can resonantly drive internal modes of
oscillation of the companion \cite{etienne29,etienne30,
etienne17, etienne31, etienne23, etienne24, etienne25, etienne,lai}, which in turn
can alter the emitted gravitational wave signal.
The mode driving can be primarily dissipative, or adiabatic, or
resonant; see for example the discussion in Ref.\ \cite{etienne}.
The modification to the emitted gravitational wave signal can
potentially yield information about the internal structure of the
stars, or impede the matched-filtering based detection of
the events.

In this paper we will focus attention on resonant mode driving.
To a good approximation, the effect of the mode resonance on the
gravitational wave signal can be described by two parameters, the
gravitational-wave frequency $f_0$ (twice the orbital frequency) at
which resonance occurs, and a constant phase shift parameter $\Delta \Phi$.
If we denote by $\Phi_0(f)$ the phase of the Fourier transform of the
gravitational waveform neglecting mode driving, then the phase $\Phi(f)$
including mode driving can be written as \cite{etienne}
\begin{equation}
\label{phaseshift} \Phi(f) = \left\{ \begin{array}{ll}
         \Phi_0(f)+(1-f/f_0)\Delta\Phi   & \mbox{$f_0-f \gg \Delta
         f_{\rm res}$} \\
         \Phi_0(f)
          & \mbox{$f-f_0 \gg \Delta f_{\rm res}.$} \\
        \end{array}
    \right.
\end{equation}
Here we have chosen the parameters of the unperturbed and perturbed
waveforms so that the waveforms coincide after the resonance.
Also $\Delta f_{\rm res}$ is the bandwidth of the resonance, which
is sufficiently small
\footnote{From
Eqs.\ (1.5) and (3.2) of Ref.\ \protect{\cite{etienne}}, the
bandwidth is
$$
\Delta f_{\rm res} \sim
0.1 \, {\rm Hz} \left(\frac{ {\cal M} }{1.2 \, M_\odot} \right)^{5/6}
\left(\frac{ f_0 }{10 \, {\rm Hz}} \right)^{11/6},
$$
where ${\cal M}$ is the chirp mass of the binary.
} that we can neglect it for data-analysis purposes.
Thus, the phase perturbation due to the resonance is a linear function
of frequency which vanishes after the resonance, whose maximum value
is $\sim \Delta \Phi$.

There are examples known of situations for which the resonant phase shift
$\Delta \Phi$ is large compared to unity, which implies the effect
should be easily detectable in the gravitational wave signal.  However
all such cases requires the spin frequency of the star to be
larger than is thought to be likely for most NS-NS inspirals.
Resonant excitation requires that the mode frequency be small compared
with the natural frequency $\omega \sim \sqrt{M/R^3}$ of the star,
where $M$ is the stellar mass and $R$ the stellar radius.
One class of modes with suitably small frequencies are $g$-modes;
however the overlap integrals for these modes are so small that
$\Delta \Phi$ is small compared to unity
\cite{etienne17,etienne29,etienne30}.
Another class are the $f$ and $p$-modes of rapidly rotating stars, where the
inertial-frame frequency $\omega_{\rm m}$ can be much smaller than the
corotating-frame frequency $\sim \sqrt{M/R^3}$.  Ho and Lai
\cite{etienne31} showed that the phase shifts due to these
modes could be large compared to unity.
However, the required NS spin frequencies are several hundred Hz,
which is thought to be unlikely in inspiralling NS-NS binaries.
A third class of modes are Rossby modes ($r$-modes)
\cite{etienne35,etienne36}, for which the
restoring force is dominated by the Coriolis force.  For these modes the
mode frequency $\omega_{\rm m}$ is of order the spin frequency of the
star, and thus can be suitably small [$10 \, \text{Hz} \lesssim \omega_\text{
  m} / (2 \pi) \lesssim 100 \, \text{Hz}$].  Ho and Lai
\cite{etienne31} computed the Newtonian driving of these
modes, and showed that the phase shift is small compared to unity.
Ref.\ \cite{etienne} showed that a larger phase shift is produced by
post-Newtonian, gravitomagnetic driving, but that once again $\Delta \Phi
\agt 1$ is only possible for spin frequencies of order several hundred Hz.

In this paper we compute how large the phase shift $\Delta \Phi$ needs
to be in order to be detectable.  We confirm the prevailing
expectation that the detectability criterion is $\Delta \Phi \agt 1$.
More specifically, we consider a simplified model of the gravitational
wave signal \cite{cutlereanna} which depends only on 5 parameters:
the masses $M_1$ and $M_2$ of the two stars, the time $t_c$ of
coalescence, the orbital phase $\phi_c$ at coalescence, and an overall
amplitude parameter ${\cal A}$ \footnote{For simplicity we neglect the
  influence of the spins of the NSs on the parameter extraction.
  Including spin will degrade somewhat the measurement accuracies.}.  Table \ref{table1}, from
Ref.\ \cite{cutlereanna}, shows the accuracy with which these parameters
could be measured in the absence of mode excitation, for an event with
signal-to-noise ratio of 10, and assuming the advanced LIGO noise
spectrum.

\begin{table}
\caption{The rms errors for signal parameters, neglecting spin effects
  and in the absence of resonances, assuming the advanced LIGO noise
  spectrum \cite{cutlereanna}, and for a signal-to-noise ratio of 10.
The units of the stellar masses $M_1$ and $M_2$
are solar masses, while $\Delta t_c$ is in milliseconds.\label{table1}}
\begin{tabular}{ll|lllll}
$M_1$&$M_2$&$\Delta t_c$&$\Delta\phi_c $&${{\Delta
M}/{M}}$&${{\Delta \mu}/{\mu}}$\\
\tableline
2.0 & 1.0 &  \ 0.72 & 1.31 & \ 0.004\% & \ 0.39\%\\
1.4 & 1.4 & \ 0.71 & 1.28 & \ 0.004\% & \ 0.41\%\\
10  & 1.4 & \ 1.01  & 1.63 & \ 0.020\% & \ 0.54\%\\
15  & 5.0 & \ 1.44  & 2.02 & \ 0.113\% & \ 1.50\%\\
10  & 10  &  \ 1.43  & 1.98 & \ 0.160\% & \ 1.90\%\\
\end{tabular}
\end{table}

Next, we enlarge the signal parameter space to include the two
parameters $f_0$, the gravitational wave frequency at
resonance, and $\Delta \Phi$, the phase shift parameter.
We compute measurement accuracies for binaries with
masses $(2
M_\odot,1 M_\odot)$, $(1.4 M_\odot,1.4 M_\odot)$, $(10 M_\odot,1.4
M_\odot)$, $(15 M_\odot,5 M_\odot)$, $(10 M_\odot,10 M_\odot)$ and
resonant frequencies $f_0=16$ Hz, $32$ Hz and
$64$ Hz with $\Delta\Phi$=1.  The results are given in
Table \ref{table2}.
The entries in the last column of this table are the minimum values of
$\Delta \Phi$ necessary for detectability of the resonance effect (see
Sec.\ \ref{sec2}).  We
see that for a $1.4 M_\odot, 1.4
M_\odot$ binary the minimum detectable values of $\Delta \Phi$ are
$\sim 8.1, 2.9$ and $1.8$ radians, for resonant frequencies of $16,
32$ and $64$ Hz.

We conclude that it is unlikely that mode resonances will be
detectable by LIGO, unless the spins of the neutron stars are
anomalously large.  Our results also show that resonances at higher
frequencies are easier to detect.


\begin{table}
\caption{The rms errors for signal parameters, as in Table
  \ref{table1}, but including the effects of resonances.  The
  resonant frequency $f_0$ is in Hz.  The quantity $\Delta (\Delta
  \Phi)$ is the rms measurement error in the phase shift parameter
  $\Delta \Phi$.\label{table2}}
\begin{tabular}{lll|llllll}
$M_1$&$M_2$&$f_0$&$\Delta{t_c}$&$\Delta \phi_c
$&${{\Delta M}/M}$&${{\Delta \mu}/
{\mu}}$&$\Delta f_0$&$\Delta(\Delta\Phi)$ \\
\tableline

2.0 & \ 1.0 & \ 16  & \ 0.81 & \ 1.68 & \ 0.007\% & \ 0.56\% & \ 24 &\ 8.1\\
1.4 & \ 1.4 & \ 16  & \ 0.80 & \ 1.65 & \ 0.008\% & \ 0.58\% & \ 24 &\ 8.1\\
10  & \ 1.4 & \ 16  & \ 1.20 & \ 2.22 & \ 0.040\% & \ 0.82\% & \ 24 &\ 8.6\\
15  & \ 5.0 & \ 16  & \ 1.82 & \ 2.89 & \ 0.232\% & \ 2.44\% & \ 24 &\ 9.2\\
10  & \ 10  & \ 16  & \ 1.81 & \ 2.85 & \ 0.318\% & \ 3.14\% & \ 24 &\ 9.2\\
2.0 & \ 1.0 & \ 32 & \ 0.76 & \ 1.46 & \ 0.007\% & \ 0.47\% & \ 17 &\ 2.9\\
1.4 & \ 1.4 & \ 32 & \ 0.75 & \ 1.43 & \ 0.007\% & \ 0.48\% & \ 17 &\ 2.9\\
10  & \ 1.4 & \ 32 & \ 1.10 & \ 1.86 & \ 0.033\% & \ 0.64\% & \ 18 &\ 2.9\\
15  & \ 5.0 & \ 32 & \ 1.64 & \ 2.36 & \ 0.165\% & \ 1.82\% & \ 18 &\ 2.9\\
10  & \ 10  & \ 32 & \ 1.61 & \ 2.32 & \ 0.224\% & \ 2.33\% & \ 18 &\ 2.9\\
2.0 & \ 1.0 & \ 64 & \ 0.97 & \ 2.44 & \ 0.005\% & \ 0.85\% & \ 23 &\ 1.8\\
1.4 & \ 1.4 & \ 64 & \ 0.95 & \ 2.39 & \ 0.005\% & \ 0.88\% & \ 23 &\ 1.8\\
10  & \ 1.4 & \ 64 & \ 1.59 & \ 3.47 & \ 0.030\% & \ 1.31\% & \ 23 &\ 2.1\\
15  & \ 5.0 & \ 64 & \ 2.48 & \ 4.55 & \ 0.202\% & \ 3.87\% & \ 23 &\ 2.5\\
10  & \ 10  & \ 64 & \ 2.45 & \ 4.45 & \ 0.280\% & \ 4.93\% & \ 23 &\ 2.5\\

\end{tabular}
\end{table}


\section{Details of analysis}
\label{sec2}

We start by reviewing the Fisher matrix formalism for computing
parameter measurement accuracies; see for example Ref.\
\cite{cutlereanna}.
The inner product used on the vector space of signals $h(t)$ is given by:
\begin{equation}
\label{inner} \left( h_1 \,|\, h_2 \right) = 2 \int_0^{\infty} \,
{ {\tilde h}_1^*(f) {\tilde h}_2(f) + {\tilde h}_1(f) {\tilde
h}_2^*(f) \over  S_n(f)} \,\,  df,
\end{equation}
where ${\tilde h}_1$ and ${\tilde h}_2$ are the Fourier transform
of two signals $h_1$ and $h_2$ respectively, and $S_n(f)$ is the
one-sided noise spectral density.
We assume the model of the advanced LIGO noise curve used in \cite{cutlereanna}
\begin{equation}
\label{snf} S_n(f) = \left\{ \begin{array}{ll}
         \infty   & \mbox{ $f < 10 \, {\rm Hz}$,} \\
         S_0 \left[(f_0/f)^4 +2 \left(1  + (f^2/f_0^2) \right) \right]
          & \mbox{    $f > 10 \,{\rm Hz}$} \\
        \end{array}
    \right.
\end{equation}
where $S_0$ is a constant whose value is unimportant for this paper,
and $f_0 = 70$ Hz.
Given a signal model $h = h(t,\theta^i)$, where $\theta^i$ are the
parameters of the signal, the Fisher information matrix is defined by
\begin{equation}
\label{sig} \Gamma_{ij} \equiv \bigg( {\partial h \over \partial
\theta^i}\, \bigg| \, {\partial h \over \partial \theta^j }\bigg),
\end{equation}
and the rms measurement error in the parameter $\theta^i$ is
\begin{equation} \label{bardx}
\sqrt{ \left< ({\Delta \theta^i})^2 \right> } =
\sqrt{\Sigma^{ii}},
\end{equation}
where ${\bf \Sigma} \equiv {\bf \Gamma}^{-1}$.

We next discuss our assumed form of the gravitational wave signal.
In the absence of resonances, we use the model of Ref.\
\cite{cutlereanna}:
\begin{equation}
\label{simplemodel}
{\tilde h}(f) ={\cal A}{f^{-7/6}}{e^{i\Phi_0(f)}},
\end{equation}
where
\begin{eqnarray}
\label{pnpsi} \Phi_0(f) & = & 2\pi f t_c -\phi_c -\pi/4 +{3\over
4}(8 \pi {\cal M} f
)^{-5/3} \nonumber \\
\mbox{} & & \times \, \left[1 + {20\over 9}\left({743\over
336}+{{11\mu}\over{4 M}}\right) x -16 \pi x^{3/2} \right].
\end{eqnarray}
Here $M$ is the total mass of the binary, $\mu$ is the reduced mass,
${\cal M}$ is the chirp
mass, and $x = (\pi M f)^{1/3}$.  We
assume the signal shuts off at a
frequency of $f = (6^{3/2}\pi M)^{-1}$.
To include the effect of resonances, we replace the phase $\Phi_0(f)$
in Eq.\ (\ref{simplemodel}) with the phase $\Phi(f)$ given by Eq.\
(\ref{phaseshift}), which depends in addition on the two parameters $f_0$ and
$\Delta \Phi$.  [We approximate the bandwidth $\Delta f_{\rm res}$ of
the resonance to be zero.]
Using this signal model given by the seven parameters
$\cal{A}$, $t_c$, $\phi_c$, $\mu$, $M$, $f_0$ and $\Delta\Phi$, we
numerically compute the Fisher matrix (\ref{sig}), and invert the
matrix to obtain the parameter measurement accuracies shown in Table
\ref{table2}.

In these computations, we use the value $\Delta \Phi=1$.  The phase
perturbation will be detectable when $\Delta \Phi \agt \Delta (\Delta \Phi)$.  From the
form of the gravitational wave signal it follows that the rms error
$\Delta (\Delta \Phi)$ is independent of the value of $\Delta \Phi$.
Therefore the minimum value of $\Delta \Phi$ necessary for detection
is given by the computed value of $\Delta (\Delta \Phi)$, i.e. the
last column in Table \ref{table2}.

\newpage

\newpage

\end{document}